\documentclass[twocolumn,trackchanges]{aastex701} 

\usepackage{siunitx}
\usepackage{dcolumn}
\usepackage{amsmath}
\DeclareSIUnit\angstrom{\textup{\AA}}
\DeclareSIUnit\mag{mag}
\DeclareSIUnit\mmag{mmag}
\DeclareSIUnit\dex{dex}
\usepackage{xcolor}
\makeatletter
\def\aas@table@head#1{%
  \noalign{\vskip 3pt}%
  \multispan\@tempcnta\hfil\bfseries #1\hfil\cr%
  \noalign{\vskip 1pt}%
}
\makeatother
\usepackage{natbib}

\renewcommand{\textbf}[1]{#1}
\begin{document}
\title{Stellar Loci X: Selection of Nitrogen-Rich Giants Using GALEX NUV and Synthetic SDSS Photometry}

\author[0009-0001-8266-3283]{Liangyu Yu}\email{202531101996@mail.bnu.edu.cn}
\affiliation{Institute for Frontiers in Astronomy and Astrophysics,
Beijing Normal University, Beijing 102206, People's Republic of China}
\affiliation{School of Physics and Astronomy, Beijing Normal University No.19, Xinjiekouwai St, Haidian District, Beijing 100875, People's Republic of China}

\author[0000-0003-2471-2363]{Haibo Yuan}\email{yuanhb@bnu.edu.cn}
\affiliation{Institute for Frontiers in Astronomy and Astrophysics,
Beijing Normal University, Beijing 102206, People's Republic of China}
\affiliation{School of Physics and Astronomy, Beijing Normal University No.19, Xinjiekouwai St, Haidian District, Beijing 100875, People's Republic of China}

\author[0000-0002-0066-0346]{Baitian Tang}\email{tangbt@mail.sysu.edu.cn}
\affiliation{School of Physics and Astronomy,
Sun Yat-sen University, Daxue Road, Zhuhai, 519082, People's Republic of China}
\affiliation{CSST Science Center for the Guangdong-Hong Kong-Macau Greater Bay Area, Zhuhai, 519082, People's Republic of China}

\author[0000-0001-8424-1079]{Kai Xiao}\email{xiaokai@ucas.ac.cn}
\affiliation{School of Astronomy and Space Science, University of Chinese Academy of Sciences, Beijing 100049, People’s Republic of China}

\author[0009-0009-0678-2537]{Xue Lu}\email{202221160014@mail.bnu.edu.cn}
\affiliation{Institute for Frontiers in Astronomy and Astrophysics,
Beijing Normal University, Beijing 102206, People's Republic of China}
\affiliation{School of Physics and Astronomy, Beijing Normal University No.19, Xinjiekouwai St, Haidian District, Beijing 100875, People's Republic of China}

\author[0000-0002-0900-9760]{Jos\'e G. Fern\'andez-Trincado}\email{jose.fernandez@ubo.cl}
\affiliation{Centro de investigaci\'on en Astronom\'ia, Facultad de Ingenier\'ia, Ciencia y Tecnolog\'ia, Universidad Bernardo O’Higgins, Av. Viel 1497, Santiago, 8370993, Chile}

\correspondingauthor{Haibo Yuan}
\email{yuanhb@bnu.edu.cn}

\begin{abstract}
Nitrogen-rich (N-rich) field stars (likely second-generation escapees from globular clusters) provide a powerful, but currently sample-limited, probe of globular-cluster disruption in the Galactic halo. We exploit the fact that enhanced NH, CN absorption depresses optical $u$ band flux, while the near-ultraviolet (NUV) band is largely insensitive to nitrogen enhancement while remaining sensitive to overall metallicity. Combining GALEX NUV photometry with (synthetic) SDSS optical photometry, we construct a $[\mathrm{Fe/H}]$-dependent stellar-locus model for giants and select significant redder $u-g$ outliers as N-rich candidates. For a relaxed cut ($[\mathrm{Fe/H}]\le-0.8$, $2\sigma$) we identify 970 candidates, \textbf{with an estimated completeness of 64--65\% and purity of 41--62\% from multiple validations}. A stricter selection ($[\mathrm{Fe/H}]\le-1.0$, $3\sigma$) yields 220 candidates, \textbf{with lower completeness (30--38\%) but substantially higher purity (60--91\%)}. The dominant contaminants show anomalous spectral energy distributions (e.g., UV excess), and are themselves useful targets for follow-up. \textbf{The diverse kinematic and dynamical properties of our candidates suggest multiple origins.} This scalable photometric approach enables systematic searches for N-rich halo stars and is well suited to forthcoming wide-field surveys such as CSST.
\end{abstract}


\section{Introduction} \label{sec:Introduction}
Spectroscopic and photometric surveys have established that globular clusters (GCs) commonly exhibit pronounced light-element abundance anomalies, most clearly seen as anti-correlations between element pairs such as N--C, Na--O and Mg--Al (e.g., \citealt{carrettaNaOAnticorrelationHB2009a,carrettaNaOAnticorrelationHB2009b,piottoHubbleSpaceTelescope2015}). These patterns are generally attributed to high-temperature proton-capture nucleosynthesis (via the CNO, NeNa and MgAl cycles; e.g., \citealt{langerSODIUMOXYGENABUNDANCEANTICORRELATIONS1993}). On this basis, GCs are typically interpreted as hosting multiple stellar populations: first-generation (FG) stars have light-element abundances comparable to field stars, whereas second-generation (SG) stars are enhanced in N, Na and Al (and possibly He and Si) and depleted in C, O and Mg (as reviewed in \citealt{miloneHubbleSpaceTelescope2017}).

The dominant theoretical framework for explaining these complex abundance patterns is the ``self-enrichment scenario'', in which SG stars formed from gas enriched by nucleosynthetic products of FG stars. Potential sources of this enrichment include intermediate-mass AGB (e.g., \citealt{dantonaSingleModelVariety2016}), fast-rotating massive stars (FRMS; \citealt{krauseSuperbubbleDynamicsGlobular2013}), interactions in massive binaries \citep{deminkMassiveBinariesSource2009}, or very massive stars (VMS; \citealt{denissenkovSupermassiveStarsSource2014}).

However, the self-enrichment scenario faces the ``mass budget problem'': the observed fraction of SG stars in GCs (40--90\%; e.g., \citealt{miloneHubbleSpaceTelescope2017}) significantly exceeding theoretical predictions (e.g., \citealt{cabrera-ziriConstrainingGlobularCluster2015}). One widely discussed solution relies on the dynamical evolution in the Milky Way gravitational field: FG stars, often spatially more extended than SG stars within GCs \citep{lardoMiningSDSSSearch2011}, are preferentially stripped due to tidal heating \citep{vesperiniFRACTIONGLOBULARCLUSTER2010}. Other escape mechanisms, such as stellar evolution and two-body relaxation in GCs, also contribute to the long-term evaporation of cluster members \citep{weatherfordStellarEscapeGlobular2023}.

The chemical fingerprints imprinted in former GC members can be preserved throughout their lifetimes. Identifying these signatures is therefore essential for testing stellar escape mechanisms and for quantifying the contribution of disrupted GCs to the Galactic halo. One of the most distinctive tracers of GC origin is the SG light-element anti-correlation pattern, for which nitrogen enhancement is often the most accessible indicator in large spectroscopic surveys; however, N-rich halo stars consistent with a GC origin are extremely rare, comprising only 2--3\% of the inner-halo population \citep{hortaContributionNrichStars2021,schiavonChemicalTaggingAPOGEE2017,belokurovNitrogenEnrichmentClustered2023,martellLightelementAbundanceVariations2010}. This scarcity has strongly limited progress in related studies.

N-rich stars can be efficiently identified in large spectroscopic surveys via the strong NH and CN molecular absorption features in the $u$ and $g$ bands. Several hundred such stars have been reported to date. Using the Sloan Digital Sky Survey (SDSS; \citealt{yorkSloanDigitalSky2000}) spectra, the CN3839 and CH band indices have been utilized to identify 118 CN-strong stars \citep{martellLightelementAbundanceVariations2010,martellBuildingGalacticHalo2011,kochPurveyorsFineHalos2019}. From the Large Sky Area Multi-Object Fiber Spectroscopic Telescope (LAMOST; \citealt{dengLAMOSTExperimentGalactic2012,zhaoLAMOSTSpectralSurvey2012}) spectra, \citet{tangChemicalKinematicAnalysis2019,tangChemicalKinematicConsistency2020} modeled the dependence of the CN3839, CN4142 and CH4300 indices on $T_{\rm eff}$ and selected 106 N-rich stars. Incomplete follow-up high-resolution spectroscopic observations (23 out of the LAMOST-selected sample) support their GC origins \citep{yuChemicalTaggingNrich2021,qiaoTaleTwoOrigins2026}. The Apache Point Observatory Galactic Evolution Experiment (APOGEE; \citealt{majewskiApachePointObservatory2017}) high-resolution near-infrared spectroscopy provides detailed abundances, enabling the identification of 412 C-depleted N-rich stars (as summarized in \citealt{fernandez-trincadoGalacticArchaeoLogIcaLExcavatiOns2022}). More recently, \citet{kaneOnesThatGot2025} trained a neural-network model that combines Gaia \citep{prustiGaiaMission2016} BP/RP (XP) spectra with APOGEE abundances, yielding 878 N-rich candidates.

Despite the success of spectroscopy in identifying N-rich stars, existing samples remain limited by observational constraints (e.g., target selection, survey depth, and pipeline complexity). Photometric approaches offer a promising route to building a larger sample with fainter limiting magnitudes and thus larger spatial coverage. 
\textbf{Pseudo two-color diagrams (``chromosome maps'') are highly sensitive to variations in C, N, O, and He abundances (e.g., \citealt{miloneHUBBLESPACEESCOPE2015}). They have been widely used to reveal multiple populations in GCs using HST UV--optical filter combinations (e.g., \citealt{piottoHubbleSpaceTelescope2015,miloneHubbleSpaceTelescope2017}) and, more recently, using pseudo-filters constructed from Gaia XP spectrophotometry \citep{mehtaSpectrophotometryRadialDistribution2025}. Although developed for cluster members, these results highlight the diagnostic power of UV--optical colors for identifying N-rich GC escapees in the field.}

\citet{zhangIdentificationNrichMetalpoor2023} assessed the ability of the Chinese Space Station Telescope (CSST; \citealt{csstcollaborationIntroductionChineseSpace2026}) main survey camera to identify N-rich field stars using broadband photometry and slitless spectroscopy. They showed that, given a metallicity estimate independent of the $u$ band, N-rich stars can be separated from normal halo stars using $u,g,r$ photometry. Motivated by this result, we propose a method that combines UV and optical photometry to select N-rich stars. Our goal is to construct a robust, large-sample catalogue and to provide a methodological foundation for future studies based on CSST data.

The physical basis for selecting N-rich stars with multi-band photometry lies in their differential sensitivity to metallicity and nitrogen abundance. Specifically, the fluxes of N-rich stars in the $u$ and $g$ bands are suppressed due to metal absorption lines and molecular absorption features associated with nitrogen, where the NH and CN bands exert a significantly stronger effect in the $u$ band than in the $g$ band. Consequently, at fixed metallicity, N-rich stars exhibit systematically redder $u-g$ colors compared to normal field stars, providing a robust photometric diagnostic for their identification. Notably, the near-ultraviolet (NUV) band is largely unaffected by NH and CN molecular absorption but remains highly sensitive to the dense forest of metal absorption lines, thereby offering a relatively independent measurement of metallicity \citep{luStellarLociVII2024}. By incorporating the NUV-based metallicity constraint, we can effectively disentangle the contributions of metallicity and nitrogen abundance to the flux suppression in the $u$ and $g$ bands, thus isolating the $u-g$ color reddening specifically attributable to nitrogen enhancement.

This paper is organized as follows. Section~\ref{sec:Data} describes the data and reduction procedures. Section~\ref{sec:Method} presents the selection method. Section~\ref{sec:Results} reports the results and validation tests. Section~\ref{sec:Discussion} 
discusses the selection efficiency and factors affecting it\textbf{, as well as the kinematic and dynamical properties of the candidates}. 
Section~\ref{sec:Conclusion} summarizes our conclusions and outlines future prospects.

\begin{figure*}[ht!]
\includegraphics[width=\textwidth]{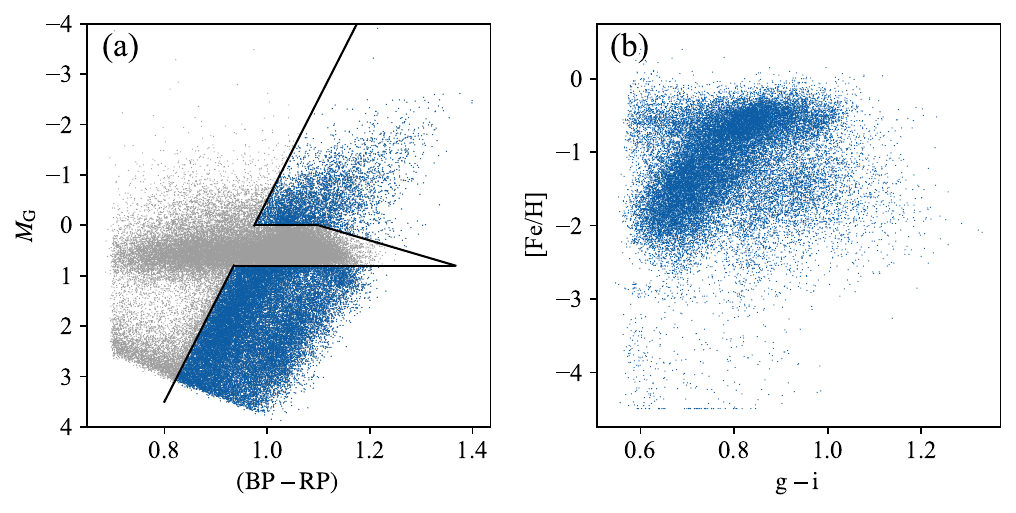}
\caption{(a) H--R diagram of the photometric sample: sources excluded by our criteria are shown in gray, and retained sources in blue; the black lines mark the HR diagram cut (criterion~3). (b) Distribution of the retained sources in the $[\mathrm{Fe/H}]$ vs. g$-$i plane.
\label{fig:sample}}
\end{figure*}

\section{Data} \label{sec:Data}

\citet{luStellarLociVII2024} combined the Galaxy Evolution Explorer (GALEX; \citealt{martinGalaxyEvolutionExplorer2005}) GR6+7 All-sky Imaging Survey (AIS; \citealt{bianchiUltravioletSkyOverview2014}) and Gaia EDR3 photometry with spectroscopic metallicities from the SAGA and PASTEL catalogues to construct a high-quality sample of $\sim$4.5 million dwarfs and $\sim$0.5 million giants. They estimated photometric metallicities using a $[\mathrm{Fe/H}]$- and $M_{\mathrm{G}}$-dependent stellar loci (NUV$-$BP versus BP$-$RP colors), achieving a typical precision of \SI{0.17}{\dex} for giants and extending down to $[\mathrm{Fe/H}]=-4.0$.

From this catalogue we select 501,358 giants, excluding stars flagged for unreliable metallicities. We then build our photometric sample by cross-matching these giants with the SDSS $ugriz$ XP synthetic photometry (XPSP; \citealt{xiaoJPLUSPhotometricRecalibration2023}) available from the BEst STar (BEST) database \citep{xiaoLargePreciseAllSky2026} with a typical precision better than \SI{5}{\mmag}.

Interstellar extinction (reddening) significantly affects UV/optical colors; therefore, we adopt the empirical reddening coefficients $R_{\rm color}(T_{\rm eff},E(B-V))$ using the \texttt{extinction-coefficient} Python package from \citet{zhangEmpiricalTemperatureExtinctiondependent2022} to apply an extinction correction. Because $T_{\rm eff}$ is not available for all sources, we use the observed BP$-$RP color as a proxy. $E(B-V)$ is obtained from the dust map of \citet{schlegelMapsDustInfrared1998}. Hereafter we assume all quoted colors are dereddened.

We then apply the following selection criteria:

1. Off-disk selection: $|Z|=(1/\varpi)\,\sin b > 1\,\mathrm{kpc}$, where $\varpi$ is corrected Gaia parallaxes from \citet{lindegrenGaiaEarlyData2021} and $b$ is Galactic latitude.

2. Photometric error: $\texttt{u\_err}<0.02$\,mag.

3. HR diagram cut: to reduce contamination from red-clump and horizontal-branch stars, we require $M_G\ge -20(BP-RP)+19.5$ and exclude stars with $0\le M_G\le 0.8$ and $M_G\ge 3(BP-RP)-3.3$ (following \citealt{huangStellarLociVIII2025}).

4. Gaia renormalised unit weight error (ruwe): to reduce non-single sources, we require $\texttt{ruwe}\le1.4$.

After applying these criteria, 32,255 sources remain. Their distributions in the HR diagram and parameter space are shown in Figure~\ref{fig:sample}, where sources to the left of the black lines are excluded by criterion~3.

\begin{figure}[ht!]
\includegraphics[width=\columnwidth]{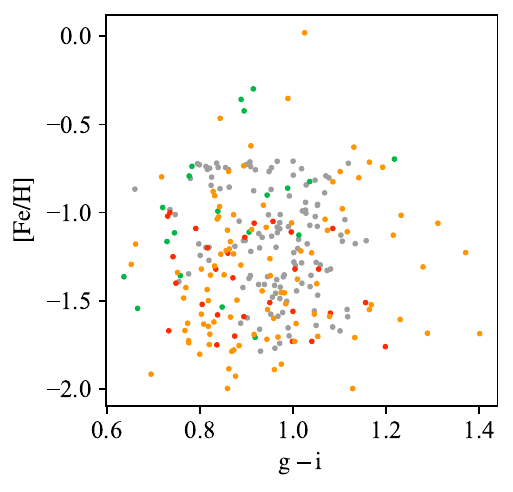}
\caption{Reference-sample distribution in the $[\mathrm{Fe/H}]$ vs. g$-$i plane: APOGEE (green), LAMOST (red), and Gaia (orange). Metallicities are adopted from Fern\'andez-Trincado et~al.\ (2022), Tang et~al.\ (2020), and Kane et~al.\ (2025), respectively. 
\textbf{We also overlay 122 APOGEE nitrogen-typical stars (gray), selected following \citet{fernandez-trincadoGalacticArchaeoLogIcaLExcavatiOns2022} and cross-matched to our photometric sample.}
\label{fig:feh}}
\end{figure}

Next, we cross-match our photometric sample with the literature N-rich stars and candidates described in Section~\ref{sec:Introduction} using a \SI{1}{\arcsecond} radius (the same hereafter). We find 1 match with \cite{kochPurveyorsFineHalos2019}, 25 with \cite{tangChemicalKinematicConsistency2020}, 33 with \cite{fernandez-trincadoGalacticArchaeoLogIcaLExcavatiOns2022}, and 107 with \cite{kaneOnesThatGot2025}. After removing duplicates and sources without synthetic photometry, 49 N-rich stars (APOGEE/LAMOST reference sample) and 99 N-rich candidates (Gaia reference sample) remain; their metallicity distributions are shown in Figure~\ref{fig:feh}.

\section{Method} \label{sec:Method}

Because N-rich stars are expected to have redder $u-g$ colors than normal stars at fixed $[\mathrm{Fe/H}]$, we adopt the $[\mathrm{Fe/H}]$-dependent stellar-locus approach of \cite{yuanStelLARLOCIMETALLICITY2015} to select them. We model the stellar locus in $u-g$ as a function of $g-i$ and $[\mathrm{Fe/H}]$ with a third-order polynomial (10 free parameters):
\begin{equation}
\begin{split}
(u - g) ={}& a_0 + a_1 X + a_2 Y + a_3 X^2 + a_4 X Y + a_5 Y^2 \\
          & + a_6 X^3 + a_7 X^2 Y + a_8 X Y^2 + a_9 Y^3,
\end{split}
\end{equation}
where $X \equiv g-i$ and $Y \equiv [\mathrm{Fe/H}]$. 
Given the absence of an explicit dependence between $u-g$ residual and $M_{\mathrm{G}}$ (related to surface gravity), we neglect the effect of $M_{\mathrm{G}}$.

We select N-rich candidates as significant high-side outliers in the $u-g$ residuals relative to the fitted locus. Most globular clusters are metal-poor (MP); therefore, we restrict the analysis to MP candidates.

A potential contaminant arises because $[\mathrm{Fe/H}]_{\mathrm{GALEX}}$, inferred from NUV photometry, can be underestimated due to UV excess (UVe). At fixed $g-i$ and $u-g$, the underestimation is degenerate with N enhancement. Consequently, normal stars exhibiting UVe may be shifted into the outlier region defined by our locus-based selection. To reduce this contamination, we apply an independent metallicity constraint to the preliminary candidate list.

Eventually, we assess the completeness and purity of the final sample using independent validation and quantify the impact of remaining contaminants (statistical scatter and UVe).

\section{Results} \label{sec:Results}

\subsection{Fitting and Selection} \label{subsec:Fitting_and_Selection}

\begin{figure*}[ht!]
\includegraphics[width=\textwidth]{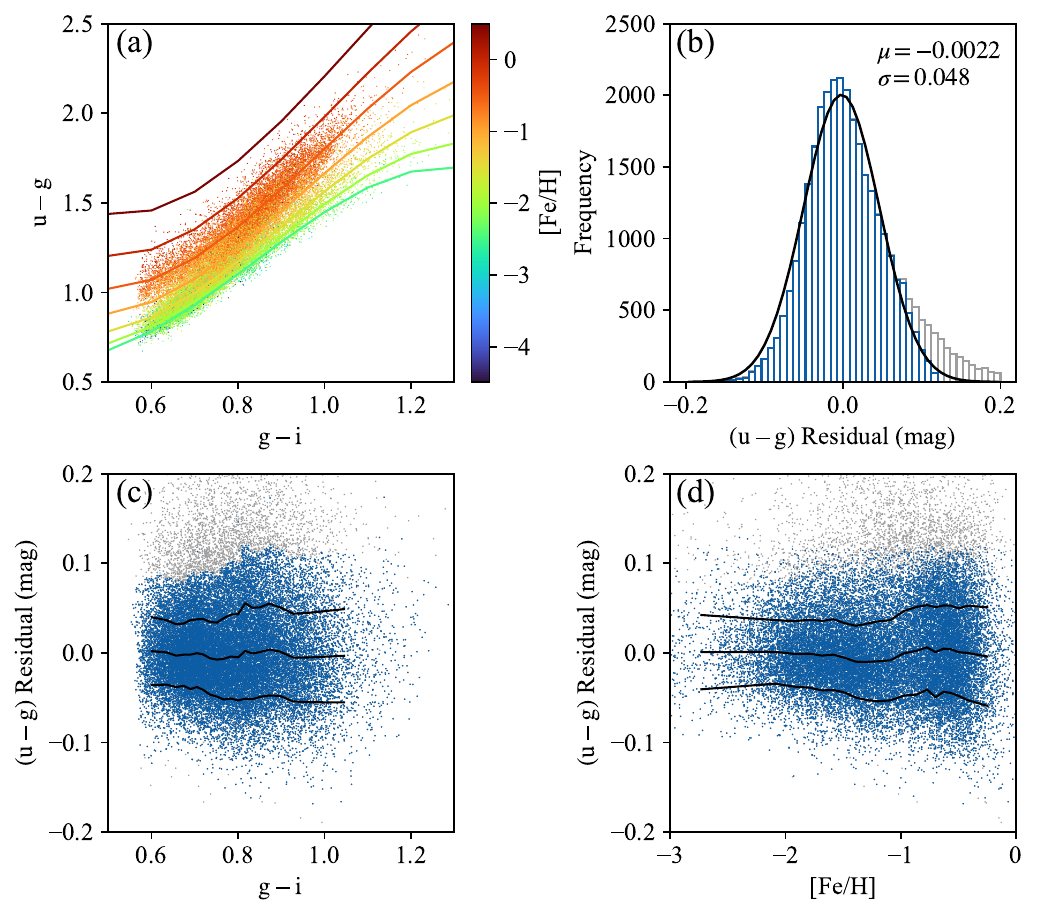}
\caption{(a) Training-sample distribution in the u$-$g vs. g$-$i plane, color-coded by $[\mathrm{Fe/H}]$; labeled curves show the fitted stellar loci at representative metallicities. (b) Histogram of $u-g$ residuals with a Gaussian fit (black). (c)--(d) $u-g$ residuals versus $g-i$ and $[\mathrm{Fe/H}]$, respectively; black curves indicate the median and $\pm1\sigma$ envelope. Points used in the fit are shown in blue, and clipped outliers in gray.
\label{fig:fit}}
\end{figure*}

Because N enhancement increases $u-g$ at fixed $[\mathrm{Fe/H}]$, N-rich stars can bias the locus fit if retained. We therefore fit with asymmetric sigma clipping (lower: $3\sigma$; upper: $2\sigma$; two iterations), remaining 29,409 sources. For these remaining sources, we model $u-g$ as a function of $g-i$ and $[\mathrm{Fe/H}]$. The best-fit coefficients are listed in Table~\ref{tab:coef}, and Figure~\ref{fig:fit} shows the resulting fit and residuals, with no significant trends.

\begin{figure}[ht!]
\includegraphics[width=\columnwidth]{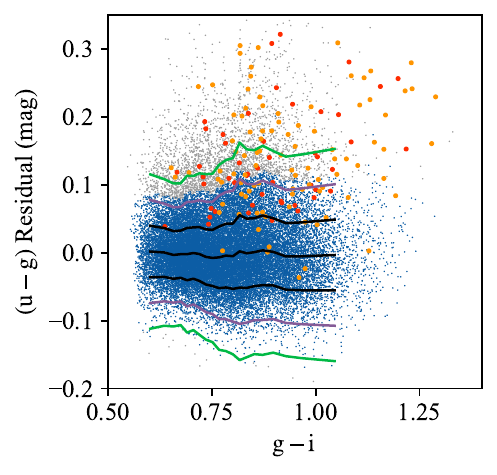}
\caption{$u-g$ residuals vs. $g-i$ for fitted sources (blue) and clipped outliers (gray), shown alongside the APOGEE/LAMOST (red) and Gaia (orange) reference samples. Black curves mark the median and $\pm1\sigma$ envelope; purple and green curves denote the $\pm2\sigma$ and $\pm3\sigma$ thresholds, respectively.
\label{fig:residuals}}
\end{figure}

Figure~\ref{fig:residuals} shows $u-g$ residuals as a function of $g-i$. The APOGEE/LAMOST and Gaia reference samples exhibit systematically high residuals, supporting the selection of N-rich candidates via high residuals. Applying a $3\sigma$ cut (more conservative) or a $2\sigma$ cut yields 1,525 and 3,660 candidates, respectively.

{
\setlength{\tabcolsep}{30pt}
\begin{deluxetable*}{cc}
\tablecaption{Stellar Locus Fitting Coefficients \label{tab:coef}}
\tablehead{\colhead{Coeff.} & \colhead{Value}}
\startdata
$a_0$ & $2.8394$ \\
$a_1$ & $-7.2007$ \\
$a_2$ & $0.7834$ \\
$a_3$ & $9.3749$ \\
$a_4$ & $-1.0895$ \\
$a_5$ & $0.1072$ \\
$a_6$ & $-3.0336$ \\
$a_7$ & $0.7089$ \\
$a_8$ & $-0.0121$ \\
$a_9$ & $0.0076$ \\
\enddata
\end{deluxetable*}
}

\subsection{$[\mathrm{Fe/H}]$ Refinement} \label{subsec:[Fe/H] Refinement}

\begin{figure}[ht!]
\includegraphics[width=\columnwidth]{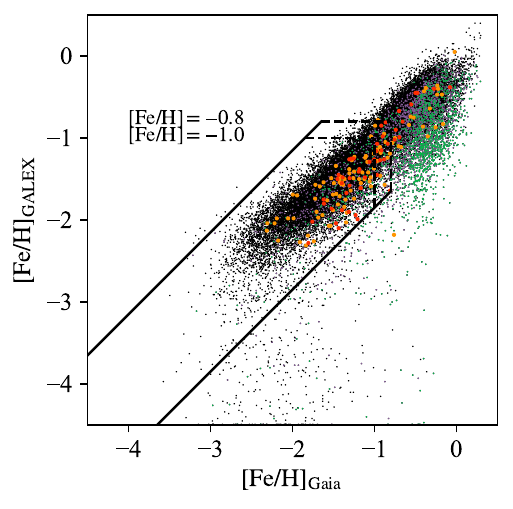}
\caption{Distribution of the photometric sample (black) in the $[\mathrm{Fe/H}]_{\mathrm {Gaia}}$--$[\mathrm{Fe/H}]_{\mathrm {GALEX}}$ plane, overlaid with the APOGEE/LAMOST (red) and Gaia (orange) reference samples, as well as the $2\sigma$ (purple) and $3\sigma$ (green) candidate selections. Solid black lines indicate $|\Delta[\mathrm{Fe/H}]|=0.85$, and dashed lines mark $[\mathrm{Fe/H}]=-0.8$ and $-1.0$.
\label{fig:refinement}}
\end{figure}

We restrict the selection to MP sources, consistent with the typical GC metallicities. However, enhanced NUV flux (i.e., UVe) can contaminate photometric metallicity estimates, biasing $[\mathrm{Fe/H}]_{\mathrm {GALEX}}$ to spuriously low values. To refine the MP selection, we adopt $[\mathrm{Fe/H}]_{\mathrm {Gaia}}$ derived from the uncertainty-aware, cost-sensitive neural network (UA-CSNet; \citealt{yangMetallicities20Million2025}) and compare it with $[\mathrm{Fe/H}]_{\mathrm {GALEX}}$ to exclude sources affected by UVe (Figure~\ref{fig:refinement}).

We define two MP subsets by requiring (i) both metallicity estimates $[\mathrm{Fe/H}]\leq Z$ and (ii) their absolute difference 
$|\Delta[\mathrm{Fe/H}]| \equiv |[\mathrm{Fe/H}]_{\mathrm {GALEX}}-[\mathrm{Fe/H}]_{\mathrm {Gaia}}| \leq 0.85\,\mathrm{dex}$.
We adopt $Z=-0.8$ and $Z=-1.0$ for the two subsets.

The $[\mathrm{Fe/H}]\leq -0.8$ subset contains 17,536 sources, including 39 APOGEE/LAMOST N-rich stars and 78 Gaia N-rich candidates. Of these, the $2\sigma$ selection yields 970 candidates 
(\textbf{25 APOGEE/LAMOST and 51 Gaia; completeness 64--65\%}), 
whereas the $3\sigma$ selection yields 291 candidates (\textbf{13 APOGEE/LAMOST and 31 Gaia; completeness 33--40\%}).
The $[\mathrm{Fe/H}]\leq -1.0$ subset contains 15,049 sources, with 27 APOGEE/LAMOST N-rich stars and 65 Gaia N-rich candidates. Of these, the $2\sigma$ selection yields 727 candidates (\textbf{15 APOGEE/LAMOST and 40 Gaia; completeness 56--62\%}), whereas the $3\sigma$ selection yields 220 candidates (\textbf{8 APOGEE/LAMOST and 25 Gaia; completeness 30--38\%}).
\textbf{We adopt $|\Delta[\mathrm{Fe/H}]|\leq 0.85\,\mathrm{dex}$ to maximize the completeness of the reference sample. Tighter cuts (0.6 and 0.4) yield essentially the same completeness and purity as 0.85, except that 0.4 lowers completeness by 5--13\% for a marginal purity gain of $<6\%$; we therefore retain 0.85 as the best compromise.}
\textbf{The quoted completeness values should be interpreted as recovery fractions of the known reference samples, which are themselves subject to selection effects (e.g., APOGEE magnitude limits, incomplete representativeness in LAMOST, and the lack of spectroscopic confirmation for Gaia XP).}

\begin{deluxetable*}{cccrccccc}
\tabletypesize{\footnotesize}
\tablewidth{\textwidth}
\tablecaption{Summary of Selection Performance \label{tab:result}}
\tablehead{
\multicolumn{2}{c}{Threshold} && \colhead{Sum} & \multicolumn{2}{c}{Completeness} && \multicolumn{2}{c}{Purity}\\
\colhead{} & \colhead{} && \colhead{} & \colhead{\shortstack{APOGEE/LAMOST}} & \colhead{Gaia} && \colhead{DD-Payne} & \colhead{LAMOST/DESI Spectra}
}
\startdata
$[\mathrm{Fe/H}] \leq -0.8$ & $2\sigma$ && 970 & 64\% (25/39) & 65\% (51/78) && 41\% (54/132) & 40-56\% (67-94/167) \\
  & $3\sigma$ && 291 & 33\% (13/39) & 40\% (31/78) && 55\% (22/40) & 63-78\% (31-38/49) \\
$[\mathrm{Fe/H}] \leq -1.0$ & $2\sigma$ && 727 & 56\% (15/27) & 62\% (40/65) && 47\% (43/91) & 49-63\% (56-73/115) \\
   & $3\sigma$ && 220 & 30\% (8/27) & 38\% (25/65) && 60\% (18/30) & 76-91\% (25-30/33) \\
\enddata
\tablecomments{%
Values are reported as ``\% ($N/M$)'': \\
Completeness $= N\,(\text{recovered reference})/M\,(\text{total reference})$, purity $= N\,(\text{classified candidates})/M\,(\text{total selected})$. \\
For spectral purity, the range corresponds to counting only grade A (lower bound) versus grades A+B (upper bound).%
}
\end{deluxetable*}

\subsection{Validation} \label{subsec:Validation}

\subsubsection{DD-Payne $[\mathrm{N/Fe}]$} \label{subsubsec:DD-Payne_[N/Fe]}

The most direct validation of our photometric selection is to examine nitrogen abundances. The Data-Driven Payne (DD-Payne; \citealt{xiangAbundanceEstimates162019}) derives stellar parameters and multi-element abundances from low-resolution spectra; recent applications include DESI EDR \citep[12 elements; $\sim5.2\times10^5$ FGK stars; most elemental abundance precision $\sim$ \SI{0.05}{\dex}]{zhangDeterminingStellarElemental2024} and LAMOST DR9 \citep[22 elements; $\sim6.4\times10^6$ FGK stars; light-element abundance precision $\sim$ \SI{0.05}{\dex}]{zhangHomogeneousStellarAtmospheric2025}.

\begin{figure*}[ht!]
\includegraphics[width=\textwidth]{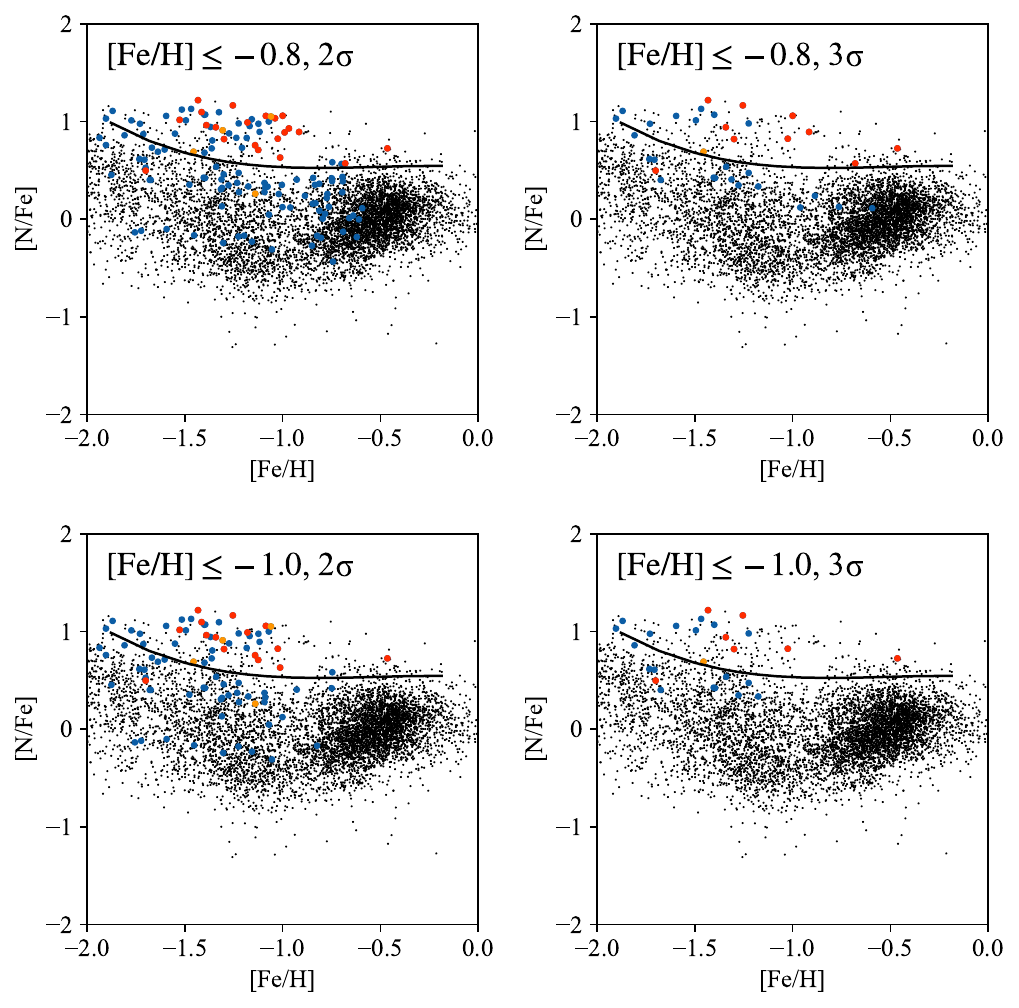}
\caption{DD-Payne $[\mathrm{N/Fe}]$--$[\mathrm{Fe/H}]$ for candidates (blue) and the full photometric sample (gray), shown alongside the APOGEE/LAMOST (red) and Gaia (orange) reference samples. The black curve shows the MWFM from \cite{fernandez-trincadoGalacticArchaeoLogIcaLExcavatiOns2022}; the selection threshold for N-rich candidates is listed in the corner of each panel.
\label{fig:abundances}}
\end{figure*}

Our photometric sample shows no overlap with the DESI EDR DD-Payne catalogue but shares 6,566 common sources with the LAMOST DR9 DD-Payne catalog after cross-matching and excluding sources with unreliable $[\mathrm{C/Fe}],[\mathrm{N/Fe}]$ measurements and $\texttt{SNR\_G} < 10$. Their $[\mathrm{Fe/H}]$ and $[\mathrm{N/Fe}]$ distributions are shown in Figure~\ref{fig:abundances}. We adopt the Milky Way Fiducial Model (MWFM) of \cite{fernandez-trincadoGalacticArchaeoLogIcaLExcavatiOns2022} to represent the mean $[\mathrm{N/Fe}]$ trend of MW normal stars and select stars above the curve to be N-rich. The purity of our candidate sample is estimated as the fraction of such objects among all candidates with available DD-Payne abundance measurements.

For the $[\mathrm{Fe/H}] \leq -0.8$ subset, 132 $2\sigma$ candidates have DD-Payne abundance and 54 are confirmed as N-rich (purity 41\%), while 40 $3\sigma$ candidates have abundance and 22 are confirmed as N-rich (purity 55\%).
For the $[\mathrm{Fe/H}] \leq -1.0$ subset, 91 $2\sigma$ candidates have abundance and 43 are confirmed as N-rich (purity 47\%), while 30 $3\sigma$ candidates have abundance and 18 are confirmed as N-rich (purity 60\%).

\begin{figure*}[ht!]
\centering
\includegraphics[width=\textwidth, height=0.9\textheight, keepaspectratio]{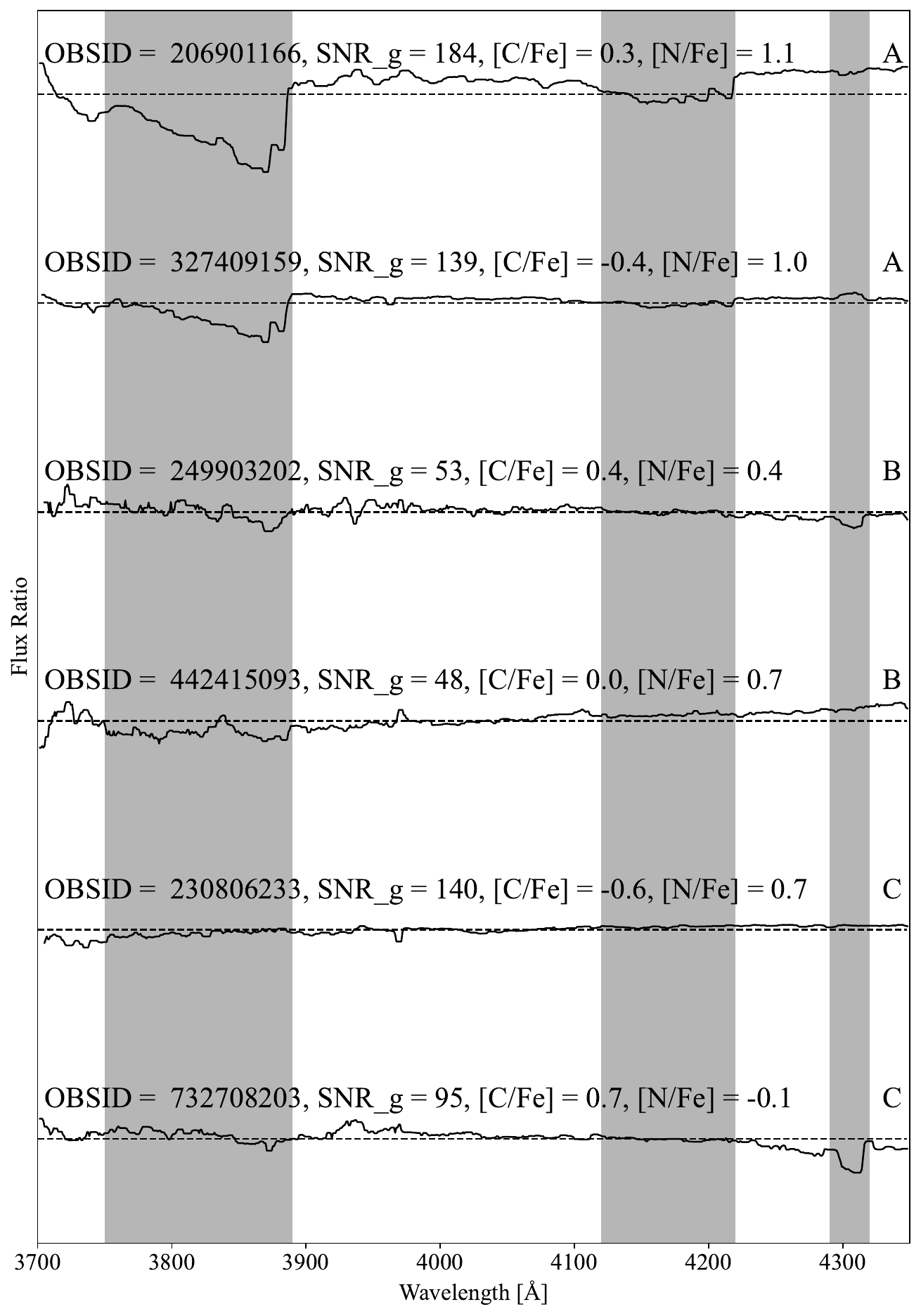}
\caption{Example flux ratio between LAMOST target and reference spectra illustrating variations in CN absorption strength. For each spectrum we annotate the \texttt{obsid}, the $g$-band SNR, the DD-Payne $[\mathrm{C/Fe}]$ and $[\mathrm{N/Fe}]$ (if available), and a qualitative grade indicating the assessor’s confidence that the star is N-rich. The CN3839, CN4142, and CH4300 molecular bands are marked by gray shaded regions. The solid curve shows the flux ratio of the candidate spectrum to its template; the dashed line marks a flux ratio of unity.
\label{fig:spectrum}}
\end{figure*}

\subsubsection{CN Absorption Bands \label{subsubsec:CN_Absorption_Bands}}

We further validate the candidates spectroscopically by comparing each target spectrum to a matched template and inspecting the flux ratio for enhanced CN absorption. For each candidate, we retrieve its reference spectra from LAMOST DR13 (low resolution) and DESI DR1 and construct a template spectrum via weighted averaging.

Reference spectra are required to have $\mathrm{SNR}\ge 10$ ($g$ band for LAMOST and $B$ arm for DESI) and to closely match the target atmospheric parameters:
$|\Delta T_{\mathrm{eff}}|\le 50\,\mathrm{K},\quad |\Delta\log g|\le 0.1\,\mathrm{dex},\quad |\Delta[\mathrm{Fe/H}]|\le 0.1\,\mathrm{dex}.$

We shift all reference spectra to the rest frame using their radial velocities. Because the corrected spectra fall on different wavelength grids, we resample them onto a common grid using linear interpolation. We then deredden the resampled spectra using \texttt{xp\_extinction\_toolkit} python package from \cite{zhangEmpiricalExtinctionCurve2024} with $E(B-V)$ from \cite{schlegelMapsDustInfrared1998}. Template spectra are computed with weights proportional to $(\mathrm{SNR})^2$. All spectra are divided by the mean flux over \SIrange{3700}{4300}{\angstrom}, and a 9-pixel median filter is applied to emphasize broad absorption features.

We then perform a visual classification based on the CN band strengths near \SI{3839}{\angstrom} and \SI{4142}{\angstrom}.  Spectra exhibiting strong CN absorption were labeled as grade A, those with weak absorption as grade B, and those without detectable CN features as grade C (see Figure~\ref{fig:spectrum}). 
\textbf{To improve robustness, three co-authors independently reviewed the visual classifications. In cases of disagreement, we adopt the most conservative (lowest) grade as the final label. All disagreements were between adjacent grades (A/B or B/C), consistent with an unambiguous separation between grades A and C.}
We estimate the purity of our candidates as the fraction of spectra showing CN absorption, using the number of grade A spectra as a conservative lower bound and the combined number of grades A+B as an upper bound.

\textbf{For the $[\mathrm{Fe/H}] \leq -0.8$ subset, the $2\sigma$ candidate sample comprises 139 LAMOST and 28 DESI spectra, with 67--94 classified as N-rich (purity 40--56\%); the $3\sigma$ sample comprises 40 LAMOST and 9 DESI spectra, with 31--38 classified as N-rich (purity 63--78\%).
For the $[\mathrm{Fe/H}] \leq -1.0$ subset, the $2\sigma$ candidate sample comprises 96 LAMOST and 19 DESI spectra, with 56--73 classified as N-rich (purity 49--63\%); the $3\sigma$ sample comprises 29 LAMOST and 4 DESI spectra, with 25--30 classified as N-rich (purity 76--91\%).}
All completeness and purity estimates are summarized in Table~\ref{tab:result}. The columns contained in the candidate catalogue are listed in Table~\ref{tab:candidates}.

\section{Discussions}\label{sec:Discussion}

\subsection{Contamination Estimate \label{subsec:Contamination_Estimate}}

If the $u-g$ residuals of normal giants were strictly Gaussian, then selecting candidates above an $+n\sigma$ threshold would inevitably include some normal stars scattered into the tail, producing statistical contamination. N-rich giants would further enhance this tail, but the expected contamination from random fluctuations can still be estimated under the Gaussian baseline.

To contextualize the \textit{validated contamination rate} from the independent checks in Section~\ref{sec:Results}, we consider two complementary estimates:
(1) \textit{theoretical contamination}: for a standard normal distribution, the one-sided probabilities of exceeding $3\sigma$ and $2\sigma$ are 0.135\% and 2.275\%, respectively, which predict the expected number of false positives;
(2) \textit{empirical contamination}: inferred from the number of confirmed normal giants below $-n\sigma$ (the left wing of the main peak), and mapped to the $+n\sigma$ tail assuming approximate symmetry of the normal-star component.

\textbf{For the $[\mathrm{Fe/H}] \leq -0.8$ subset, the 970 $2\sigma$ candidates have a validated contamination rate of 44\%--60\%, compared with theoretical and empirical expectations of $399/970 = 41\%$ and $213/970 = 22\%$, respectively. The 291 $3\sigma$ candidates have a validated contamination rate of 22\%--45\%, compared with $24/291 = 8\%$ (theoretical) and $12/291 = 4\%$ (empirical).
For the $[\mathrm{Fe/H}] \leq -1.0$ subset, the 727 $2\sigma$ candidates have a validated contamination rate of 37\%--53\%, compared with $342/727 = 47\%$ (theoretical) and $166/727 = 23\%$ (empirical). The 220 $3\sigma$ candidates have a validated contamination rate of 9\%--40\%, compared with $20/220 = 9\%$ (theoretical) and $9/220 = 4\%$ (empirical).}

Above results indicate that at the $2\sigma$ threshold, statistical scatter can account for a substantial fraction of the contamination. In contrast, at the $3\sigma$ threshold, the residual contamination is more likely to reflect intrinsic anomalies in the sample, such as anomalous spectral energy distributions (SEDs) that induce biases in photometric metallicity estimates.

\subsection{UV excess \label{subsec:UV_Excess}}

Figure~\ref{fig:refinement} shows that $[\mathrm{Fe/H}]_{\mathrm{GALEX}}$ is lower than $[\mathrm{Fe/H}]_{\mathrm{Gaia}}$ for a portion of sources. A plausible contributor is that UVe biases the NUV-based metallicity estimate.

The far-ultraviolet (FUV)$-$NUV color provides a diagnostic for UVe. For red giants, the photospheric contribution in the GALEX FUV band is expected to be negligible, so FUV detections typically trace non-photospheric emission (e.g., hot companions such as DA white dwarfs or sdB stars, unresolved binaries, or chromospheric activity in active systems; \citealt{smithINTERESTINGFEATURESCOMBINED2014}).

\begin{figure}[ht!]
\includegraphics[width=\columnwidth]{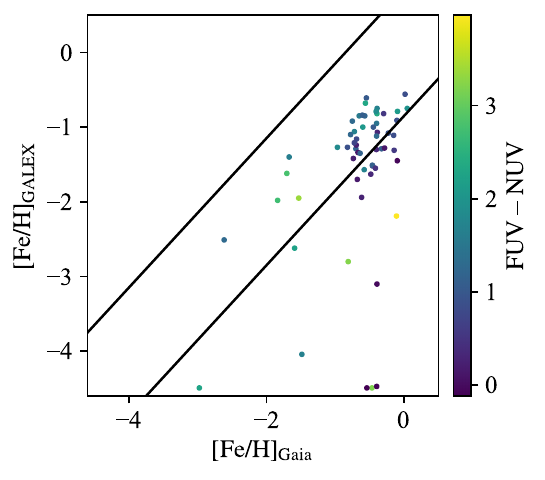}
\caption{Photometric sample in the $[\mathrm{Fe/H}]_{\mathrm {Gaia}}$--$[\mathrm{Fe/H}]_{\mathrm {GALEX}}$ plane, color-coded by FUV$-$NUV. Solid black lines indicate $|\Delta[\mathrm{Fe/H}]|=0.85\,\mathrm{dex}$.
\label{fig:fuv}}
\end{figure}

In our photometric sample, 270 sources have GALEX FUV detections; 58 have high-reliability measurements with $\mathrm{e\_FUVmag}\le 0.3$~mag. Their FUV$-$NUV distribution is shown in Figure~\ref{fig:fuv}. Some candidates lying between the black lines exhibit slightly bluer FUV$-$NUV colors. Due to their modest UVe, the resulting discrepancy between $[\mathrm{Fe/H}]_{\mathrm {GALEX}}$ and $[\mathrm{Fe/H}]_{\mathrm {Gaia}}$ may be sufficient to satisfy the $\Delta[\mathrm{Fe/H}] \le 0.85$ criterion, allowing them to remain in the sample. Such remaining UVe sources may constitute an important (possibly dominant) contaminant in the candidate catalogue.

\begin{deluxetable*}{clc}
\tablewidth{\textwidth}
\tablecaption{Candidate Catalogue Columns \label{tab:candidates}}
\tablehead{\colhead{Field} & \colhead{Description} & \colhead{Unit}}
\startdata
source\_id & Unique Gaia EDR3 source ID & \dots \\
ra & R.A. from Gaia EDR3 & deg \\
dec & Decl. from Gaia EDR3 & deg \\
pmra & Proper motion in right ascension (Gaia EDR3) & $\mathrm{mas\, yr^{-1}}$ \\
pmra\_error & Uncertainty in proper motion in right ascension (Gaia EDR3) & $\mathrm{mas\, yr^{-1}}$ \\
pmdec & Proper motion in declination (Gaia EDR3) & $\mathrm{mas\, yr^{-1}}$ \\
pmdec\_error & Uncertainty in proper motion in declination (Gaia EDR3) & $\mathrm{mas\, yr^{-1}}$ \\
rpgeo & Median photogeometric distance (posterior; \cite{bailer-jonesEstimatingDistancesParallaxes2021}) & pc \\
b\_rpgeo & 16th-percentile (lower) photogeometric distance from \cite{bailer-jonesEstimatingDistancesParallaxes2021} & pc \\
B\_rpgeo & 84th-percentile (upper) photogeometric distance from \cite{bailer-jonesEstimatingDistancesParallaxes2021} & pc \\
rv & Radial velocity & $\mathrm{km \, s^{-1}}$ \\
e\_rv & Uncertainty in radial velocity & $\mathrm{km\, s^{-1}}$ \\
flag\_rv & Radial-velocity source flag (GALAH, APOGEE, Gaia, LAMOST\_MRS, LAMOST\_LRS, SEGUE) & \dots \\
uSDSS\_xpsp & XPSP SDSS \ensuremath{u} magnitude from \cite{xiaoLargePreciseAllSky2026} & \dots \\
gSDSS\_xpsp & XPSP SDSS \ensuremath{g} magnitude from \cite{xiaoLargePreciseAllSky2026} & \dots \\
rSDSS\_xpsp & XPSP SDSS \ensuremath{r} magnitude from \cite{xiaoLargePreciseAllSky2026} & \dots \\
iSDSS\_xpsp & XPSP SDSS \ensuremath{i} magnitude from \cite{xiaoLargePreciseAllSky2026} & \dots \\
zSDSS\_xpsp & XPSP SDSS \ensuremath{z} magnitude from \cite{xiaoLargePreciseAllSky2026} & \dots \\
u\_g0 & Intrinsic \ensuremath{(u-g)_0} color & \dots \\
g\_i0 & Intrinsic \ensuremath{(g-i)_0} color & \dots \\
feh\_GALEX & Photometric metallicity from \cite{luStellarLociVII2024} & \dots \\
feh\_Gaia & Metallicity from \cite{yangMetallicities20Million2025} & \dots \\
n\_sigma & $n-\sigma$ threshold: 2 or 3 & \dots \\
cut\_feh & $[\mathrm{Fe/H}]$ refinement threshold: -0.8 ($[\mathrm{Fe/H}] \leq -0.8$), -1.0 ($[\mathrm{Fe/H}] \leq -1.0$) & \dots \\
FEH & Metallicity from \cite{zhangHomogeneousStellarAtmospheric2025}, if available & \dots \\
C\_FE & Carbon abundance from \cite{zhangHomogeneousStellarAtmospheric2025}, if available & \dots \\
N\_FE & Nitrogen abundance from \cite{zhangHomogeneousStellarAtmospheric2025}, if available & \dots \\
obsid & Unique LAMOST spectra ID, if available & \dots \\
CN\_LAMOST & CN band strength grade of LAMOST spectra: \texttt{A}, \texttt{B} or \texttt{C}, if available & \dots \\
TARGETID & Unique DESI target ID, if available & \dots \\
CN\_DESI & CN band strength grade of DESI spectra : \texttt{A}, \texttt{B} or \texttt{C}, if available & \dots \\
Vr & Galactocentric radial velocity (if computed) & $\mathrm{km\,s^{-1}}$ \\
Vphi & Galactocentric azimuthal velocity (if computed) & $\mathrm{km\,s^{-1}}$ \\
perigalacticon & Perigalactocentric distance (if computed) & $\mathrm{kpc}$ \\
Error\_perigalacticon & Uncertainty in perigalactocentric distance (if computed) & $\mathrm{kpc}$ \\
apogalacticon & Apogalactocentric distance (if computed) & $\mathrm{kpc}$ \\
Error\_apogalacticon & Uncertainty in apogalactocentric distance (if computed) & $\mathrm{kpc}$ \\
eccentricity & Orbital eccentricity (if computed) & \dots \\
Error\_eccentricity & Uncertainty in orbital eccentricity (if computed) & \dots \\
Zmax & Maximum vertical excursion from the Galactic plane (if computed) & $\mathrm{kpc}$ \\
Error\_Zmax & Uncertainty in $Z_{\max}$ (if computed) & $\mathrm{kpc}$ \\
Lzmin & Minimum $z$-component of specific angular momentum (if computed) & $\mathrm{km\,kpc\,s^{-1}}$ \\
Lzmax & Maximum $z$-component of specific angular momentum (if computed) & $\mathrm{km\,kpc\,s^{-1}}$ \\
Emean & Mean specific orbital energy (if computed) & $\mathrm{km^2\,s^{-2}}$ \\
Error\_Emean & Uncertainty in mean specific orbital energy (if computed) & $\mathrm{km^2\,s^{-2}}$ \\
\enddata
\end{deluxetable*}

\subsection{Orbital analysis \label{subsec:Orbital_analysis}}

We used the \texttt{GravPot16} Milky Way potential\footnote{\url{https://gravpot.utinam.cnrs.fr}} to integrate the orbits of the selected candidates in a steady-state Galactic model that includes perturbations from a rotating ``boxy/peanut" bar, consistent with current structural and dynamical constraints for the Milky Way.

For the orbit computations, we adopted the same model configuration, solar position, and velocity vector as in \citet{Fernandez-Trincado2019}, except for the bar pattern speed ($\Omega_{\rm bar}$), which we set to 41 km s$^{-1}$ kpc$^{-1}$ \citep{Sanders2019} and varied by $\pm$10 km s$^{-1}$ kpc$^{-1}$. 



We estimated the most likely orbital parameters and their uncertainties using a Monte Carlo approach. Specifically, we integrated $5\times 10^{5}$ orbit realizations backward for 2.5~Gyr, perturbing the observed inputs (positions, heliocentric distances, radial velocities, and proper motions) by drawing $1\sigma$ Gaussian deviations.

We computed orbits for 786 of the 970 N-rich candidates with complete six-dimensional phase-space information (Table~\ref{tab:candidates}). Radial velocities were adopted from \citet{luStellarLociVII2024}. Proper motions and their uncertainties were taken from Gaia EDR3 \citep{Gaia2021}. Heliocentric distances ($d_{\odot}$) and uncertainties were adopted from the 16th, 50th, and 84th percentiles of the Bayesian distance estimates of \citet{bailer-jonesEstimatingDistancesParallaxes2021}. The resulting orbital parameters are reported in Table~\ref{tab:candidates}.

Among the 786 candidates, 314 (40\%) exhibit disk-like kinematics with $V_{\phi}>-100\,\mathrm{km\,s^{-1}}$, whereas 472 (60\%) show Gaia--Enceladus--Sausage-like \citep[GES-like;][]{belokurovCoformationDiscStellar2018,helmiMergerThatLed2018} or halo-like kinematics with $V_{\phi}\le-100\,\mathrm{km\,s^{-1}}$. These fractions are broadly consistent with the 34\% disk-like and 65\% GES-like/halo-like proportions reported by \citet{fernandez-trincadoGalacticArchaeoLogIcaLExcavatiOns2022}.

Based on the orbital orientation relative to the Galactic-bar rotation, we divide the sample into three classes: 583 candidates (74\%) are prograde ($L_{z,\min}\le 0$ and $L_{z,\max}\le 0$) with $V_{\phi}\lesssim -10\,\mathrm{km\,s^{-1}}$; 126 (16\%) are retrograde ($L_{z,\min}\ge 0$ and $L_{z,\max}\ge 0$) with $V_{\phi}> -10\,\mathrm{km\,s^{-1}}$; and 77 (10\%) have prograde--retrograde (P--R) orbits ($L_{z,\min}<0$ and $L_{z,\max}>0$) with $-40\,\mathrm{km\,s^{-1}}\lesssim V_{\phi}<30\,\mathrm{km\,s^{-1}}$, typically associated with chaotic or strongly perturbed orbital configurations.

\section{Conclusion} \label{sec:Conclusion}

Motivated by the redder $u-g$ colors expected for N-rich giants at fixed $[\mathrm{Fe/H}]$, and $[\mathrm{Fe/H}]$ derived from GALEX NUV unaffected by N enhancement, we developed a photometric selection method that combines GALEX NUV data with XPSP SDSS $ugriz$ photometry. Using a $[\mathrm{Fe/H}]$-dependent stellar-locus model, we efficiently identify N-rich giant candidates and substantially expand the searchable sample relative to purely spectroscopic approaches.

\textbf{For the $[\mathrm{Fe/H}] \leq -0.8$ subset, the $2\sigma$ selection yields 970 candidates with a completeness of 64--65\% and a purity of 40--56\%, whereas the $3\sigma$ selection yields 291 candidates with a completeness of 33--40\% and a purity of 55--78\%.
For the $[\mathrm{Fe/H}] \leq -1.0$ subset, the $2\sigma$ selection yields 727 candidates with a completeness of 56--62\% and a purity of 47--63\%, whereas the $3\sigma$ selection yields 220 candidates with a completeness of 30--38\% and a purity of 60--91\%.}
Overall, tightening the selection decreases completeness but increases purity, illustrating the standard completeness--purity trade-off.

Notably, we find that $[\mathrm{Fe/H}]_{\mathrm{GALEX}}$ can be biased low in the presence of UVe. Incorporating an independent and more robust metallicity indicator is therefore essential for mitigating contamination. Many of these apparent ``contaminants'' are themselves astrophysically interesting, underscoring the broader potential of our method to identify stars with anomalous SEDs.

\textbf{In addition, the diversity of kinematic and dynamical properties of our candidates suggests multiple origins.}

Finally, we will apply our pipeline to the upcoming CSST main survey. With its depth, wide field of view, and photometric precision, CSST should enable the construction of a large and homogeneous N-rich candidate sample, and the brighter part of the sample will be further refined with low-resolution slitless spectra. 
\textbf{CSST is planned to include a new $v$-band filter covering the Ca\,II H\&K lines, enabling metallicity estimates that are less sensitive to UV excess. CSST will also deliver multi-epoch NUV photometry ($\geq 4$ epochs), allowing us to flag and exclude active stars and thereby reduce the false-positive rate among N-rich candidates.}
This sample would provide strong constraints on GC chemical evolution, stellar escape mechanisms and the assembly of the Galactic halo.

\begin{acknowledgments}

The authors thank the anonymous referee for suggestions that improved the clarity of our presentation. This work is supported by the National Key Basic R\&D Program of China via 2024YFA1611601 and 2024YFA1611901, the National Natural Science Foundation of China through the projects NSFC 12222301, 12173007, and 12403024.
This work is also supported by the China Manned Space Program with grant no. CMS-CSST-2025-A13.
B. Tang gratefully acknowledges support from the National Natural Science Foundation of China through grants NOs. 12233013 and 12473035.
J.G.F-T gratefully acknowledges the support provided by ANID Fondecyt Regular No.1260371.
This work made use of the data from LAMOST (Large Sky Area Multi-Object Fiber Spectroscopic Telescope, also known as the Guoshoujing Telescope) (https://cstr.cn/31118.02.LAMOST ). LAMOST is a Chinese national mega-science facility, operated by National Astronomical Observatories, Chinese Academy of Sciences.
This work has made use of data from the European Space Agency (ESA) mission {\it Gaia} (\url{https://www.cosmos.esa.int/gaia}), processed by the Gaia Data Processing and Analysis Consortium (DPAC, \url{https:// www.cosmos.esa.int/web/gaia/dpac/ consortium}). Funding for the DPAC has been provided by national institutions, in particular, the institutions participating in the Gaia Multilateral Agreement.
\end{acknowledgments}

\bibliography{sample701}
\bibliographystyle{aasjournalv7}

\end{document}